\PassOptionsToPackage{unicode}{hyperref}
\PassOptionsToPackage{hyphens}{url}
\documentclass[
]{article}
\usepackage{amsmath,amssymb}
\usepackage{iftex}
\ifPDFTeX
  \usepackage[T1]{fontenc}
  \usepackage[utf8]{inputenc}
  \usepackage{textcomp} 
\else 
  \usepackage{unicode-math} 
  \defaultfontfeatures{Scale=MatchLowercase}
  \defaultfontfeatures[\rmfamily]{Ligatures=TeX,Scale=1}
\fi
\usepackage{lmodern}
\ifPDFTeX\else
\fi
\IfFileExists{upquote.sty}{\usepackage{upquote}}{}
\IfFileExists{microtype.sty}{
  \usepackage[]{microtype}
  \UseMicrotypeSet[protrusion]{basicmath} 
}{}
\makeatletter
\@ifundefined{KOMAClassName}{
  \IfFileExists{parskip.sty}{%
    \usepackage{parskip}
  }{
    \setlength{\parindent}{0pt}
    \setlength{\parskip}{6pt plus 2pt minus 1pt}}
}{
  \KOMAoptions{parskip=half}}
\makeatother
\usepackage{xcolor}
\usepackage[margin=1in]{geometry}
\usepackage{color}
\usepackage{fancyvrb}

\DefineVerbatimEnvironment{Highlighting}{Verbatim}{commandchars=\\\{\}}
\usepackage{framed}
\definecolor{shadecolor}{RGB}{248,248,248}
\newenvironment{Shaded}{\begin{snugshade}}{\end{snugshade}}

\newcommand{\AttributeTok}[1]{\textcolor[rgb]{0.13,0.29,0.53}{#1}}

\newcommand{\CommentTok}[1]{\textcolor[rgb]{0.56,0.35,0.01}{\textit{#1}}}

\newcommand{\ConstantTok}[1]{\textcolor[rgb]{0.56,0.35,0.01}{#1}}

\newcommand{\DecValTok}[1]{\textcolor[rgb]{0.00,0.00,0.81}{#1}}

\newcommand{\FloatTok}[1]{\textcolor[rgb]{0.00,0.00,0.81}{#1}}
\newcommand{\FunctionTok}[1]{\textcolor[rgb]{0.13,0.29,0.53}{\textbf{#1}}}

\newcommand{\NormalTok}[1]{#1}

\newcommand{\OtherTok}[1]{\textcolor[rgb]{0.56,0.35,0.01}{#1}}

\newcommand{\SpecialCharTok}[1]{\textcolor[rgb]{0.81,0.36,0.00}{\textbf{#1}}}

\newcommand{\StringTok}[1]{\textcolor[rgb]{0.31,0.60,0.02}{#1}}

\usepackage{graphicx}
\makeatletter
\def\maxwidth{\ifdim\Gin@nat@width>\linewidth\linewidth\else\Gin@nat@width\fi}
\def\maxheight{\ifdim\Gin@nat@height>\textheight\textheight\else\Gin@nat@height\fi}
\makeatother
\setkeys{Gin}{width=\maxwidth,height=\maxheight,keepaspectratio}
\makeatletter
\def\fps@figure{htbp}
\makeatother
\ifLuaTeX
  \usepackage{selnolig}  
\fi
\usepackage{bookmark}
\IfFileExists{xurl.sty}{\usepackage{xurl}}{} 
\hypersetup{
  pdftitle={Treatment heterogeneity with right-censored outcomes using grf},
  pdfauthor={Erik Sverdrup; Stefan Wager},
  hidelinks,
  pdfcreator={LaTeX via pandoc}}

\title{Treatment heterogeneity with right-censored outcomes using
\texttt{grf}}
\author{Erik Sverdrup \and Stefan Wager}
\date{Stanford GSB, February 2024}

\begin{document}
\maketitle

This article walks through how to estimate conditional average treatment
effects (CATEs) with right-censored time-to-event outcomes using the
function \texttt{causal\_survival\_forest} (Cui et al., 2023) in the
\texttt{R} package \texttt{grf} (Athey et al., 2019, Tibshirani et al.,
2024). We consider a setting where we observe an unconfounded binary
treatment \(W_i \in \{0, 1\}\), and are interested in estimating its
effect on a restricted mean survival time (RMST) conditional on
covariates \(X_i\),

\[
\tau(x) = E[\textrm{min}(T_i(1),~ h) - \textrm{min}(T_i(0),~ h) ~|~ X_i = x],
\] where \(T_i(1), T_i(0)\) are potential survival outcomes in the two
treatment arms, and \(h\) is some chosen maximum considered time. Since
we are in a time-to-event setting, we don't necessarily observe the
survival time \(T_i\); instead, we observe the recorded time
\(Y_i = \textrm{min}(T_i, C_i)\) where \(C_i\) is the censoring time for
the \(i\)-th unit, along with the event indicator
\(D_i = 1(T_i \leq C_i)\). We assume that censoring is ignorable; see
Cui et al.~(2023) for a precise statement.

The function \texttt{causal\_survival\_forest} estimates CATEs using a
flexible forest-based approach that extends the random forest algorithm
of Breiman (2001). The approach starts by estimating a number of
auxiliary ``nuisance'' components, such as propensity scores and
survival curves for the survival and censoring process, using regression
forests and random survival forests respectively. The approach then
trains a final forest that directly targets CATEs via a
Neyman-orthogonal construction using these nuisance component estimates.
For complete details on the algorithm, we refer to Cui et al.~(2023).

\textbf{Remark.} \texttt{causal\_survival\_forest} also supports
estimating the effect on survival probabilities conditional on
covariates. To target the quantity

\[
\tau(x) = P[T_i(1) > h ~|~ X_i=x] - P[T_i(0) > h ~|~ X_i = x],
\] one simply uses the argument
\texttt{target\ =\ "survival.probability"} when calling into
\texttt{causal\_survival\_forest}.

\subsection{Data}\label{data}

We use data from the National Job Training Partnership Act (JPTA) study
that assessed the impact of eligibility for an employment training
program on employment outcomes using a randomized controlled trial. All
participants were interviewed at enrollment before randomization; they
were then re-interviewed sometime between one and three years
post-enrollment. The study enrolled participants who were unemployed at
the time, and one natural question to ask using data from this study is:
How did being randomized into treatment affect the amount of time it
took for study participants to find a job (Crommen et al., 2023,
Frandsen, 2015)? This time-to-event outcome is subject to random
censoring because not all participants had yet found a new job at the
time of the second survey (which was administered with different delays
post-enrollment for different participants); however, the censoring
appears plausibly ignorable here and so causal survival forests can be
used to correct for it.

The data set we're employing is a processed copy of the JPTA data made
available on GitHub by Crommen et al.~(2023), which contains
approximately 11 000 entries where subjects record the outcome (days of
unemployment), an event indicator (where 1 denotes the subject found a
job and 0 denotes censoring), a treatment indicator, as well as a
handful of covariates such as age, high school education, race,
children, marital status, and gender.

\begin{Shaded}
\begin{Highlighting}[]
\CommentTok{\# Read in publicly available JTPA data from Crommen et al. (2023).}
\NormalTok{url }\OtherTok{=} \FunctionTok{paste}\NormalTok{(}
  \StringTok{"https://raw.githubusercontent.com/GillesCrommen/DCC/"}\NormalTok{,}
  \StringTok{"748bd7f98feccad09205ee3df76df5ba740cc3d7/"}\NormalTok{,}
  \StringTok{"clean\_dataset\_JTPA.csv"}\NormalTok{, }\AttributeTok{sep =} \StringTok{""}\NormalTok{)}
\NormalTok{data }\OtherTok{=} \FunctionTok{read.csv}\NormalTok{(url)}

\CommentTok{\# Outcome (days of unemployment observed)}
\NormalTok{Y }\OtherTok{=}\NormalTok{ data}\SpecialCharTok{$}\NormalTok{days}
\CommentTok{\# Treatment indicator (W = 1 if the subject was eligible to enroll in JTPA)}
\NormalTok{W }\OtherTok{=}\NormalTok{ data}\SpecialCharTok{$}\NormalTok{treatment}
\CommentTok{\# Non{-}censoring indicator (D = 1 means the subject had a job by the second survey)}
\NormalTok{D }\OtherTok{=}\NormalTok{ data}\SpecialCharTok{$}\NormalTok{delta}
\CommentTok{\# Covariates}
\NormalTok{X }\OtherTok{=} \FunctionTok{cbind}\NormalTok{(}
  \AttributeTok{age =}\NormalTok{ data}\SpecialCharTok{$}\NormalTok{age,}
  \AttributeTok{high.school.diploma =}\NormalTok{ data}\SpecialCharTok{$}\NormalTok{hsged,}
  \AttributeTok{race.white =}\NormalTok{ data}\SpecialCharTok{$}\NormalTok{white,}
  \AttributeTok{children =}\NormalTok{ data}\SpecialCharTok{$}\NormalTok{children,}
  \AttributeTok{married =}\NormalTok{ data}\SpecialCharTok{$}\NormalTok{married,}
  \AttributeTok{male =}\NormalTok{ data}\SpecialCharTok{$}\NormalTok{male}
\NormalTok{)}
\end{Highlighting}
\end{Shaded}

Our estimand is the RMST, and so we need to choose some suitable
truncation time for the mean survival time to be identified. Based on
the histogram of outcomes by event status, we set the truncation time to
2 years (\(h=720\) days), at which point most subsequent observations
are censored.

\begin{Shaded}
\begin{Highlighting}[]
\FunctionTok{hist}\NormalTok{(Y[D}\SpecialCharTok{==}\DecValTok{1}\NormalTok{], }\AttributeTok{main =} \StringTok{"Histogram of Y"}\NormalTok{,  }\AttributeTok{xlab =} \StringTok{""}\NormalTok{)}
\FunctionTok{hist}\NormalTok{(Y[D }\SpecialCharTok{==} \DecValTok{0}\NormalTok{], }\AttributeTok{col =} \FunctionTok{adjustcolor}\NormalTok{(}\StringTok{"red"}\NormalTok{, }\FloatTok{0.5}\NormalTok{), }\AttributeTok{add =} \ConstantTok{TRUE}\NormalTok{)}
\FunctionTok{legend}\NormalTok{(}\StringTok{"topright"}\NormalTok{, }\FunctionTok{c}\NormalTok{(}\StringTok{"Event"}\NormalTok{, }\StringTok{"Censored"}\NormalTok{),}
       \AttributeTok{col =} \FunctionTok{c}\NormalTok{(}\StringTok{"gray"}\NormalTok{, }\FunctionTok{adjustcolor}\NormalTok{(}\StringTok{"red"}\NormalTok{, }\FloatTok{0.5}\NormalTok{)),}
       \AttributeTok{lwd =} \DecValTok{4}\NormalTok{)}
\FunctionTok{abline}\NormalTok{(}\AttributeTok{v =} \DecValTok{720}\NormalTok{, }\AttributeTok{lty =} \DecValTok{2}\NormalTok{)}
\end{Highlighting}
\end{Shaded}

\includegraphics{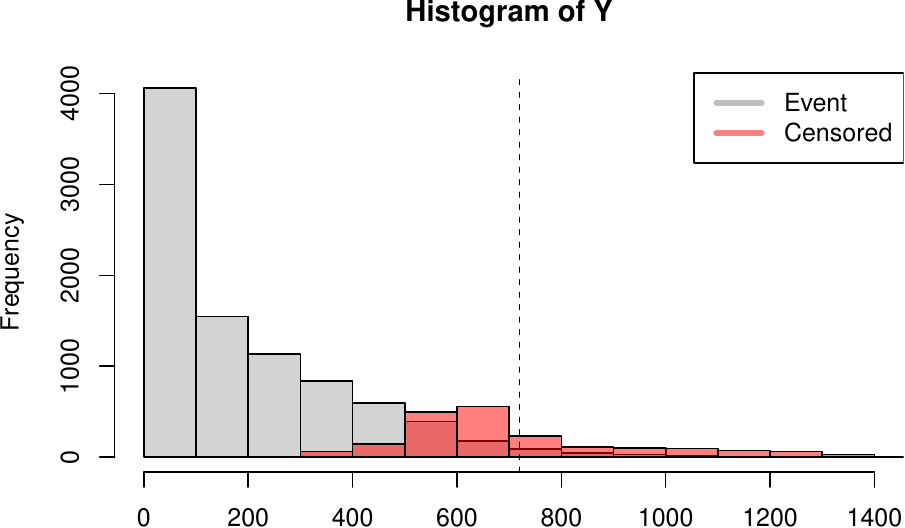}

\subsection{Fitting a causal survival forest and assessing
heterogeneity}\label{fitting-a-causal-survival-forest-and-assessing-heterogeneity}

Since the trial was randomized, we are analyzing the data as an RCT by
passing the sample average treatment fraction as the propensity score
via the argument \texttt{W.hat}. The truncation parameter \(h\) is
supplied with the argument \texttt{horizon}. Below we fit a causal
survival forest with default options and calculate a doubly robust
estimate for the average treatment effect (ATE)
\(E[\textrm{min}(T_i(1),~ h) - \textrm{min}(T_i(0),~ h)]\).

\textbf{\emph{Note.}} To avoid potential confusion with a negative sign
representing a beneficial treatment effect, we relabel the treatment
indicator such that we are measuring the effect of not being offered the
program, which will have a positive sign if the training program is
beneficial.

\begin{Shaded}
\begin{Highlighting}[]
\FunctionTok{library}\NormalTok{(grf)}

\CommentTok{\# Relabel treatment to measure the effect of not being offered training}
\NormalTok{W.relabeled }\OtherTok{=} \DecValTok{1} \SpecialCharTok{{-}}\NormalTok{ W}

\NormalTok{cs.forest }\OtherTok{=} \FunctionTok{causal\_survival\_forest}\NormalTok{(}
\NormalTok{  X, Y, W.relabeled, D,}
  \AttributeTok{W.hat =} \FunctionTok{mean}\NormalTok{(W.relabeled),}
  \AttributeTok{horizon =} \DecValTok{720}\NormalTok{)}

\CommentTok{\# Get a doubly robust estimate for the ATE}
\FunctionTok{average\_treatment\_effect}\NormalTok{(cs.forest)}
\end{Highlighting}
\end{Shaded}

\begin{verbatim}
## estimate  std.err 
##     27.2      5.4
\end{verbatim}

The ATE is positive and significant, indicating that not being offered
the program led to a longer spell of unemployment. That is, the program
appears on average to be beneficial.

\begin{Shaded}
\begin{Highlighting}[]
\CommentTok{\# Retrieve out{-}of{-}bag CATE estimates}
\NormalTok{tau.hat }\OtherTok{=} \FunctionTok{predict}\NormalTok{(cs.forest)}\SpecialCharTok{$}\NormalTok{predictions}

\FunctionTok{summary}\NormalTok{(tau.hat)}
\end{Highlighting}
\end{Shaded}

\begin{verbatim}
##    Min. 1st Qu.  Median    Mean 3rd Qu.    Max. 
##     -80       0      24      27      51     163
\end{verbatim}

If we look at the distribution of the CATE estimates we see that most
point estimates for \(\hat \tau(X_i)\) are large and positive,
suggesting the program offering reduces the unemployment duration, and
that the magnitude of the effect varies across individuals.

The numbers above are noisy non-parametric point estimates, and by
themselves don't necessarily yield insight into robust signs of
treatment effect heterogeneity. \texttt{grf} comes with various
functionality for analyzing heterogeneity based on what questions we are
interested in answering. We can conduct inference on a
``pseudo''-parameter of the CATE with the function
\texttt{best\_linear\_projection} which gives a doubly robust estimate
of the linear model \(\tau(X_i) = \beta X_i + \varepsilon_i\), where
\(X_i\) are some chosen covariates.

\begin{Shaded}
\begin{Highlighting}[]
\FunctionTok{best\_linear\_projection}\NormalTok{(cs.forest, X)}
\end{Highlighting}
\end{Shaded}

\begin{verbatim}
## 
## Best linear projection of the conditional average treatment effect.
## Confidence intervals are cluster- and heteroskedasticity-robust (HC3):
## 
##                     Estimate Std. Error t value Pr(>|t|)  
## (Intercept)           44.359     19.380    2.29    0.022 *
## age                   -0.623      0.563   -1.11    0.268  
## high.school.diploma    4.118     10.803    0.38    0.703  
## race.white           -15.096     11.069   -1.36    0.173  
## children               9.532     12.613    0.76    0.450  
## married               29.184     14.620    2.00    0.046 *
## male                  -9.282     11.975   -0.78    0.438  
## ---
## Signif. codes:  0 '***' 0.001 '**' 0.01 '*' 0.05 '.' 0.1 ' ' 1
\end{verbatim}

This analysis suggests that treatment heterogeneity is largely explained
by marital status here (married participants benefit more); however, we
caution that the statistical significance is not strong enough to
withstand a multiple testing correction (e.g., a Bonferroni correction).

To get a single p-value with power to test whether there is any
treatment heterogeneity, we can use the function
\texttt{rank\_average\_treatment\_effect} (RATE). This function tests
whether the \emph{estimated} CATE function \(\hat \tau(\cdot)\) can find
stable subgroups that benefit more from the treatment than others. This
is operationalized using the \emph{Targeting Operator Characteristic}
(TOC) curve, which for each fraction \(q\), compares the average
treatment effect of the top q-th fraction of units as prioritized by a
rule \(S(X_i)\) to the overall average treatment effect. \[
\textrm{TOC}(q) = E[\textrm{min}(T_i(1),~ h) - \textrm{min}(T_i(0),~ h) \,|\, S(X_i) \geq \textrm{q-th quantile}] - \textrm{ATE}
\] If we let \(S(X_i)\) equal the estimated CATEs,
\(S(X_i) = \hat \tau(X_i)\), we get a TOC that quantifies the difference
in ATE from the overall ATE among units predicted to have a large
increase in unemployment duration.

\begin{Shaded}
\begin{Highlighting}[]
\CommentTok{\# Form a doubly robust estimate of RATE using cs.forest}
\NormalTok{rate }\OtherTok{=} \FunctionTok{rank\_average\_treatment\_effect}\NormalTok{(cs.forest, tau.hat)}

\CommentTok{\# Plot the TOC, along with 95 \% confidence bars}
\FunctionTok{plot}\NormalTok{(rate,}
     \AttributeTok{main =} \StringTok{"TOC: By decreasing CATE estimates"}
\NormalTok{)}
\end{Highlighting}
\end{Shaded}

\includegraphics{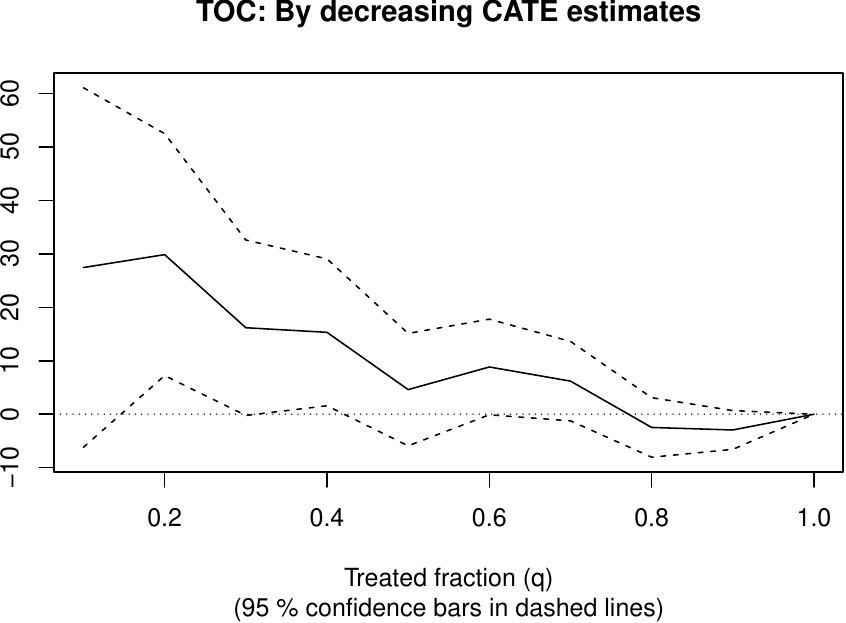}

The plot indicates there are signs of heterogeneity, for example, the
\(q=20\) percent of units with the largest estimated effect (as
predicted by \(\hat \tau\)) have an unemployment duration that is around
on average 30 days longer than the overall ATE of 27 days.

We can translate this visual stratification exercise into a point
estimate by forming an estimate of the area under this curve. A 95 \%
confidence interval for the area under the TOC does not cover 0, which
suggests that causal forests are in fact detecting real treatment
heterogeneity here:

\begin{Shaded}
\begin{Highlighting}[]
\FunctionTok{paste}\NormalTok{(}\StringTok{"AUTOC:"}\NormalTok{, }\FunctionTok{round}\NormalTok{(rate}\SpecialCharTok{$}\NormalTok{estimate, }\DecValTok{2}\NormalTok{), }\StringTok{"+/{-}"}\NormalTok{, }\FunctionTok{round}\NormalTok{(}\FloatTok{1.96} \SpecialCharTok{*}\NormalTok{ rate}\SpecialCharTok{$}\NormalTok{std.err, }\DecValTok{2}\NormalTok{))}
\end{Highlighting}
\end{Shaded}

\begin{verbatim}
## [1] "AUTOC: 13.87 +/- 11.58"
\end{verbatim}

Given an estimated function \(\hat \tau(\cdot)\) that appears to detect
meaningful heterogeneity, there are a variety of ways we can probe
further into how this heterogeneity can be described using available
covariates \(X_i\). One simple option is to examine the distributions of
covariates across subregions defined in terms of quantiles of the
estimated CATEs. Here we simply conclude with a brief example of showing
the covariate means for the full sample, and the covariate mean for the
20\% of units with the largest predicted decrease in unemployment
duration, which appears to indicate there are a larger fraction of
parents and married people in the group that is predicted to have a
large effect.

\begin{Shaded}
\begin{Highlighting}[]
\NormalTok{q80 }\OtherTok{=} \FunctionTok{quantile}\NormalTok{(tau.hat, }\FloatTok{0.8}\NormalTok{)}

\FunctionTok{rbind}\NormalTok{(}
  \AttributeTok{full.sample =} \FunctionTok{apply}\NormalTok{(X, }\DecValTok{2}\NormalTok{, mean),}
  \AttributeTok{top.20 =} \FunctionTok{apply}\NormalTok{(X[tau.hat }\SpecialCharTok{\textgreater{}=}\NormalTok{ q80, ], }\DecValTok{2}\NormalTok{, mean)}
\NormalTok{)}
\end{Highlighting}
\end{Shaded}

\begin{verbatim}
##             age high.school.diploma race.white children married male
## full.sample  29                0.48       0.56     0.49    0.23 0.43
## top.20       27                0.45       0.50     0.63    0.38 0.41
\end{verbatim}

For more details on the RATE metric, as well as strategies to avoid
over-fitting based on train/test splits, see Yadlowsky et al.~(2021), as
well as the online package documentation at
\url{https://grf-labs.github.io/grf/}.

\subsection{References}\label{references}

Susan Athey, Julie Tibshirani, and Stefan Wager. ``Generalized random
forests.'' \emph{The Annals of Statistics}, 47(2), 2019.

Leo Breiman. ``Random forests.'' \emph{Machine learning}, 45, 2001.

Gilles Crommen, Jad Beyhum, and Ingrid van Keilegom. ``An instrumental
variable approach under dependent censoring.'' \emph{TEST}, 2023.

Yifan Cui, Michael R. Kosorok, Erik Sverdrup, Stefan Wager, and Ruoqing
Zhu. ``Estimating heterogeneous treatment effects with right-censored
data via causal survival forests.'' \emph{Journal of the Royal
Statistical Society: Series B}, 85(2), 2023.

Brigham R. Frandsen. ``Treatment effects with censoring and
endogeneity.'' \emph{Journal of the American Statistical Association},
110(512), 2015.

Julie Tibshirani, Susan Athey, Rina Friedberg, Vitor Hadad, David
Hirshberg, Luke Miner, Erik Sverdrup, Stefan Wager, and Marvin Wright.
``grf: Generalized random forests.'', 2024. URL
\url{https://github.com/grf-labs/grf}. R package version 2.3.2.

Steve Yadlowsky, Scott Fleming, Nigam Shah, Emma Brunskill, and Stefan
Wager. ``Evaluating treatment prioritization rules via rank-weighted
average treatment effects.'' \emph{arXiv preprint arXiv:2111.07966},
2021.

\end{document}